\documentclass{an}
\usepackage{graphicx}
\usepackage{times}
\Volume{00}              
\Year{0000}              
\Month{00}               
\Pagespan{000}{000}      

\begin{document}
\lhead[\thepage]{A.N. Covone and Napolitano: 
Kinematics of intermediate luminosity 
early-type galaxies}
\rhead[Astron. Nachr./AN~{\bf XXX} (200X) X]{\thepage}
\headnote{Astron. Nachr./AN {\bf 32X} (200X) X, XXX--XXX}

\title{Toward a detailed view on the kinematics of intermediate luminosity 
early-type galaxies no dark matter candidates}

\author{G. Covone\inst{1} and N. R. Napolitano\inst{2}}
\institute{Laboratoire d'Astrophysique de Marseille,
traverse du Siphon, 13012 Marseille, France
\and
Kapteyn Astronomical Institute, Landeleven 12, Post Office Box 800, 
9700 AV Groningen, the Netherlands
}
\date{Received {date will be inserted by the editor}; 
accepted {date will be inserted by the editor}}

\abstract{
In several nearby $L \sim L_*$ early-type galaxies, 
recent observations at large radii have shown a indications of a lack 
of dark matter, substantially at odds with 
the  prediction from the Cold Dark Matter
(CDM)  hierarchical merger models.
Here we discuss a pilot observational project 
for the study of the internal kinematical and
dynamical properties of this remarkable sample of galaxies.
Using the VIMOS-IFU in its high spectral resolution mode,
it would be possible to 
investigate the regions up to $\sim ~1.2 ~R_{\rm e}$,
taking advantage of the much larger field of view and 
telescope diameter. This will allow to disclose 
the presence of any kinematical substructures which could affect the 
conclusion on the mass modeling and definitely clarify the inner structure 
of this particular class of early-type galaxies.
\keywords{galaxies: elliptical;
galaxies: kinematics and dynamics;
instrumentation: spectrographs}
}\correspondence{giovanni.covone@oamp.fr}

\maketitle

\section{Introduction}
Intermediate luminosity early-type galaxies ($L\sim L_*$) 
are interesting laboratories where to test both the galaxy formation theories and the presence of the dark matter.
Their structural and kinematical properties make them a different class 
with respect to the brightest systems. Indeed, from the photometric point 
of view, bright galaxies have usually boxy isophotes and flat inner cores 
in the light distribution while fainter galaxies are mostly disky with 
power-law inner light profiles (Nieto \& Bender 1989, Capaccioli et al. 1992,
Faber et al. 1998).\\
These isophote shapes appear to be correlated with a variety 
of properties of  these systems
(Kissler-Patig 1997, Pellegrini 1999 and references therein),
in particular with the kinematics: 
bright/boxy systems are slowly rotating in the inner parts, while 
faint/disky galaxies are rotationally supported at least within $1~R_{\rm e}$
(the effective radius $R_{\rm e}$ encloses half of the galaxy 
projected light), 
the typical observational limit for kinematical studies by integrated light 
long-slit spectroscopy (Bertola \& Capaccioli 1975, 
Davies et al. 1983, Scorza \& Bender 1995).

Recently, some lines of evidence has been found about the persistence 
of such a dichotomy also in the dark matter properties of these systems.
Capaccioli et al. (2003) and Napolitano et al. (2003) have discussed the 
correlation between the structural parameters and the radial trend in the
mass-to-light ratios (M/L) using archive kinematic data up to 
$\sim 6 R_{\rm e}$ for an heterogeneous sample of ellipticals: 
faint/disky galaxies 
show small or null M/L radial gradients, i.e., almost constant 
M/L, while bright/boxy objects 
have larger gradients in agreement with a substantial 
amount of dark matter in the outskirts of those systems. In particular the 
evidence of a dearth of dark matter in the intermediate luminosity galaxies 
has been confirmed by Romanowsky et al. (2003) using planetary nebulae
(PNe) kinematics out to 4 $R_{\rm e}$. 
Confirmation of these preliminary results
by new observations with larger PNe samples or new techniques
is very important, 
since a possibly tight connection between structural properties and 
density environment (i.e., the dark matter halo)
where the galaxies have grown would deeply affect our understanding of galaxy formation and evolution.

The formation and evolution of these intermediate 
luminosity galaxies with almost
null M/L gradients is still not clear.
As a class, disky galaxies can be considered close to the bulge dominated 
S0 galaxies since disky isophotes could  reflect the 
presence of an embedded faint stellar disk (Capaccioli 1987; Bender et al. 
1988). 
Scorza \& Bender (1995) have shown that the information 
on both photometry and line profile of a sample of disky 
galaxies is compatible with simple disk+bulge models where the two 
subcomponents have parallel angular momenta. This latter evidence could be 
an hint of a contemporary evolution for these galaxies rather than a late 
accretion or merger events.\\ 
However, disky isophotes are also expected in hierarchical merging 
scenarios in a CDM cosmology (Meza et al. 2003): boxy/disky isophotes are found for inclinations which minimize/maximize $V_{\mathrm{rot}}/\sigma$.
Furthermore, clues on angular momentum alignment could change when looking at the halo regions of these systems. 
Preliminary results on PNe radial velocity fields of NGC~3379 and 
NGC~4494 from the Planetary Nebula Spectrograph, (PN.S, Douglas et al. 2002), 
have revealed
that there is a tilt of the major kinematical 
axis of the PNe sample with respect to the stellar long-slit data as it 
is clearly shown in Fig. 1. 
In the same figure a quite good alignment is found for NGC821 as well as it has been found also in earlier studies (NGC~4697: Mend\`ez et al. 2001). 
Incidentally, the latter are 
viewed quite edge-on, while NGC~3379 and NGC~4494 are possibly 
nearly face-on. 
Furthermore, at least for the four systems mentioned above, 
the rotation 
velocity curves increase in the inner parts (up to $\sim 
1~R_{\rm e}$) and rapidly drop in the outer parts 
down to zero at the outermost points.
Is this an indication of some secondary evolution processes in some of 
these galaxies (i.e., tidal interactions or merging), 
or is this the result of different stellar population inhabiting inner (disk population) and outer (bulge/halo population) 
regions, which appear to dominate depending on the viewing angle? 

The geometry can play a major role even in modeling the mass distribution, 
in particular for the systems seen almost face--on
(Magorrian \& Ballantyne 2001).
If the  presence of a disk is kinematically relevant up to large radii, then we 
could expect radial orbits to dominate in projection, causing an apparent 
decreasing velocity dispersion and then a mass underestimate, i.e., 
we would erroneously conclude that there is a lack of dark matter.

So far traditional long-slit spectroscopy has been employed to 
address these problems, resulting in one dimensional information 
mostly performed along the main principal photometrical/kinematical axes 
of galaxies (Bender, Saglia \& Gerhard 1994, Statler \& Smecker-Hane 1999).
Nowadays, integral field unit spectrographs are opening a new era in the 
understanding of the galaxy structure and evolution.
We discuss here an observational project to study a preliminary 
sample of intermediate luminosity/disky galaxies using the 
Integral Field Unit (IFU) of VIMOS.
The most important wide field 
integral field spectroscopy instruments 
used to study nearby elliptical galaxies
are shortly described in Sect. 2, guideline of the project are 
briefly described in Sect. 3, together with the main expected
results.

\begin{figure*}
\resizebox{\hsize}{!}
{\includegraphics[]{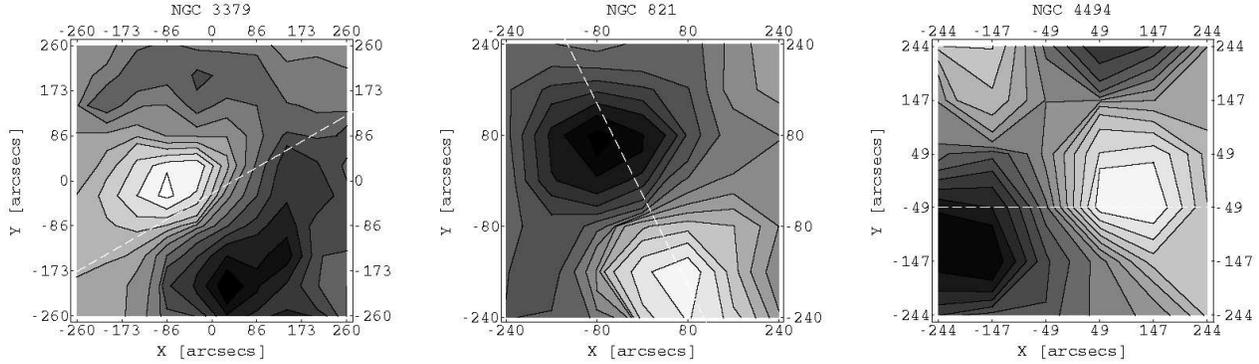}}
\caption{Gaussian smoothed PNe velocity field from PN.S data (R+03), 
darker to brighter regions are negative to positive radial velocities. 
Dashed line is the P.A. of the inner stellar major kinematical axis. 
Misalignment of the PNe axis with respect the inner stellar kinematics, 
kinematical substructures and twist of the isovelocity contours are evident.
Signatures of such features in the inner stellar population can be 
investigated by VIMOS-IFU in great detail.}
\label{figlabel}
\end{figure*}

\section{Integral field spectroscopy of nearby galaxies}

Since integral field spectrographs are available on 
4-meter class telescopes, 
the study of the kinematics and the internal structure of local 
early-type galaxies has
been one of the most important applications of such instruments. 
In particular, a long term project has been undertaken by the 
SAURON team (de Zeeuw et al. 2002),
using the integral field instrument SAURON (Bacon et al. 2001)
at the WHT (La Palma).
SAURON has been the first IFU with a 
field of view (f.o.v.) larger than $\sim 10''$, i.e.,
wide enough to obtain data up to $0.5 R_{\rm e}$ in a 
single pointing for nearby luminous galaxies.

The SAURON project is a large and ambitious survey of a representative 
sample of 72 nearby early-type (E, S0 and Sa) galaxies ($z \leq 0.01$).
The f.o.v.
is $33'' \times 41''$ in the low resolution mode 
($\Delta \lambda=3.6 $ \AA, i.e., $R \sim 1400$, each fiber covering
$0''.94$), and $9'' \times 11''$ in the high resolution mode
($\Delta \lambda= 2.8 $ \AA, $R \sim 1800$ at 5000 \AA, each fiber 
covering $0''.28$). 
The 1520 spectra cover the range from 4800 to 5400 \AA.
The instrumental velocity dispersion $\sigma_{\rm inst}$
is 105 and 90 km s$^{-1}$, in the two
modes, allowing to measure the line of sight velocity dispersion of stellar
spheroidal components  of galaxies that have velocity dispersions larger than
about 100 km s$^{-1}$.
First results from this SAURON survey of 
early-type galaxies are published in de Zeeuw et al. (2002).

At the moment a new IFU is available to the astronomical community,
with a wider
field of view, the VIMOS - IFU.
It is one of the three possible operation modes of VIMOS,
the new survey-designed instrument mounted on VLT-UT3: wide field imaging,
multi-object spectroscopy (up to several hundred slits)
and integral field spectroscopy.
Detailed information on the instrument, the 
IFU and its operational modes 
can be found at the ESO web site\footnote{\tt
http://www.eso.org/instrument/vimos};
see also Foucaud et al. (2003) for a brief description of the
data analysis and D'Odorico et al. (2003) for 
a review of its performance.
Up to now, the VIMOS-IFU has the largest f.o.v.
among the instrument devoted to integral field spectroscopy: up to 
$54'' \times 54''$ in the low (spatial and spectral) resolution mode,
allowing to obtain up to 6400 spectra. In high spatial resolution 
mode the f.o.v. is $27'' \times 27''$ and fibers of $0''.67$ 
can be used for high spectral resolution.

The HR-Blue grism (420-615 nm)
covers the spectral range traditionally 
used to study kinematics of local galaxies with a spectral resolution, 
$R \sim 2500$, i.e. $\Delta \lambda = 2.0 $ \AA \,
at $\lambda = 5000$ \AA; 
indeed it allows to observe the following spectral lines in the local Universe:

\begin{itemize}
\item Mg b triplet @$\lambda$5170 \AA 
\item Fe lines @$\lambda\lambda$5270, 5335 \AA 
\item $\left[ {\rm O III } \right]$ @$\lambda\lambda$4959, 5007 \AA
\item H$\beta$, @$\lambda$4861 \AA 
\end{itemize}
while with the redder grism HR-Red (635-860 nm)
it will be possible to cover also the H$\alpha$
emission line. 
The first two of these lines 
would allow to probe the stellar kinematics, while 
$\left[ {\rm O III } \right]$
and  H$\beta$ can be used to derive information on the properties of 
the ionized gas (morphology, kinematics and ionization state).
The larger wavelength coverage, compared to the one used 
from the SAURON team 
will allow to pursue the same kinematical and
dynamical study beyond the local Universe.
\footnote{In the near future, it is planned 
that SAURON will operate also in the range 4500- 7000 \AA, 
in order to cover also the H$\alpha$ emission line,
but for the same set of $z <  0.01$ galaxies.}

There are some other advantages of VIMOS-IFU with respect to SAURON: 
in the higher resolution mode the field of view is comparable to 
the one of SAURON, but with a slightly lower instrumental dispersion
($\sigma_{\rm inst} \simeq 80 $  km s$^{-1}$ at $\lambda \simeq 5000 $ \AA),
allowing to study galaxies with
slightly lower velocity dispersion, and the smaller spatial elements
dimensions ($0''.67$) is well suited to the excellent 
average seeing on Paranal.
Of course, VIMOS takes advantage of the VLT collecting power: 
for example,  
using the HR-Blue grism, combined with the high spatial resolution mode, 
it will be possible to 
to cover the region within 1.2 $R_{\rm e}$ of NGC 3779 with 6 different
pointings in 2 nights (see next section).
In general, it will be possible to reach the same signal-to-noise ratio as in
SAURON observations in about half of
the exposure time (since, taking into account background effects, 
the exposure time scales as the inverse of the diameter).

\section{Deep IFS of no dark matter candidate}

\begin{table*}[]
\caption{Intermediate luminosity galaxies
no dark matter candidates.}
\begin{center}
\begin{tabular}{lllllllll} 
\hline
galaxy & Type & $V_{sys} $ & $R_{\rm e} $ & $M_B$ & a$_4$& 
$\gamma$ & $\Gamma_{B5} $ & reference \\
& &km s$^{-1}$ & & & & & $(\Gamma_{B \odot})$ & \\
\hline 
\hline 
 NGC821  & E2     &  1735 & $50''$ & -20.5 & 2.5  & 0.64 & 13-17    & R+03\\
 NGC3779 & E1/S0? &  877  & $35''$ & -20.0 & 0.2 & 0.18 & 5-8      & R+03\\ 
 NGC4494 & E0     &  1232 & $49''$ & -20.6 & 0.3  & 0.6  & 5-7      & R+03\\ 
 NGC4697 & E3     &  1310 & $90''$ & -20.5 & 1.4  & 0.74 & 12$\pm$1 & Np01\\
\hline
\end{tabular}
\end{center}
\label{table:galaxies}
\end{table*} 

In Table~\ref{table:galaxies} we show the characteristics
of a sample of galaxies for which previous observations at large
radii have shown the intriguing result of 
presence of little, if any, dark matter.  

These galaxies have the common properties to 
have disky isophotes (a$_4> $ 0) and a power-law slope of the inner density 
profile ($\gamma > $ 0), while their 
M/L at 5 effective radii, $\Gamma_{B5}$, obtained with 
extended PNe kinematics\footnote{For NGC4697, 
Napolitano (2001) obtained a constant M/L integrating the Jeans 
equations up to infinity.} 
are in good agreement
with typical stellar mass-to-light ratios.
This suggests a small amount 
of dark matter enveloping the luminous part of these systems.
It is interesting to note that the possibility that 
many ordinary early-type galaxies are deficient in dark matter 
content was already discussed in previous dynamical studies (e.g., Bertin et al.
1994).
Since this result clearly  does not conform with 
the CDM evolution scenarios, 
more detailed observations are needed to confirm it.
\footnote{In this respect, it is interesting to note that the
MOND theory appears to be in agreement with the findings
of Romanowsky et al. (Milgrom \& Sanders 2003).}

Such a kind of studies would be greatly advantaged by 
a full 2D kinematical map of such a galaxies, 
as it could be obtained by combining the 2D 
discrete radial velocity field of PNe with the integral field data. 
In Fig. 1 a smoothed version of the PNe radial velocity fields Romanowsky et al. (2003) is shown: this is the kind of data which we expect to complement with the VIMOS-IFU observations. In particular we expect to obtain: 

\begin{itemize}
\item a description of the fine structure of the stellar kinematics  
with a spatial resolution of $0''.67$, mapping the line-of-sight 
rotation velocity $V$, velocity dispersion $\sigma_{\rm v}$, and higher 
Gauss-Hermite coefficients $h_3$, $h_4$. 
These maps can be quantified via Fourier methods. 

\item the possibility to reveal the presence of subcomponents along the 
line-of-sight both from $h_3$ and $h_4$ distribution and possibly using 
the double-Gaussian fit as performed in Scorza \& Bender 
(1995)\footnote{Here they investigate the internal structure of disky 
systems with long-slit spectroscopy and instrumental errors similar to 
those expected with IFU-VIMOS.}.    

\item the association between kinematical and photometrical structures 
(using reconstructed continuum image from IFU data) as well as better 
constraints on the intrinsic inclination (see, e.g., Verolme et al. 2002).

\end{itemize}

For instance, with 6 VIMOS-IFU pointings it is possible to cover an area 
of 81$''\times 54''$ around NGC3379 center, corresponding to 
1.2$ R_{\rm e}\times 0.8 R_{\rm e}$ 
with an average S/N of 30 in a total of 14 hours integration. 
In this area a sample of at least 60 PNe is present, 
which is expected to be sufficient for any check of consistency between 
kinematical estimates (Napolitano et al. 2001).

Using such a full 2D kinematical map, we expect to definitely disclose the 
presence of any 
substructure that could affect the projected kinematics like faint disks 
seen at high inclination, by means of the azimuthal modulation of the 
velocity moments on the sky plane (for instance, a non-cylindrical 
rotation velocity in proximity of the main kinematical axis).

For the galaxies in Table \ref{table:galaxies},
Romanowsky et al. (2003) have applied a full spherical 
Schwarzshild orbit library method (Romanowsky \& Kochanek 2001) 
for modeling the PNe kinematics with different degree of anisotropy. 
Such a modeling approach 
could be greatly improved with the 2D information of 
the stellar kinematics in the inner parts, where 
higher order velocity moments can help to break the 
mass-anisotropy degeneracy (Lokas 2002). 
The results of such a modeling 
procedure can be compared with the use of 2 (or 3) integral Jeans models for 
axisymmetric systems (Pignatelli \& Galletta 1999, Napolitano et al. 2001).

\section{Conclusions}

The ubiquity of dark matter in early-type galaxies is a debated topic:
already Saglia et al. (1993), using stellar kinematics observations,
hinted at the possible existence of two classes of
early-type galaxies, according to the velocity dispersion trend at large radii.
They studied several early-type galaxies with decreasing 
velocity dispersion which required no dark
matter (see also Capaccioli et al. 2003).
However, long-slit spectroscopical stellar kinematics studies are 
generally confined to the inner luminous regions ($R < R_{\rm e}$).
A remarkable improvement has been possible using PNe as
kinematical tracers 
in the outskirts of
early-type galaxies with dedicated instruments like the PN.S:
preliminary results have already been
published, and a larger sample of early-type galaxies is 
being observed with the same technique.
However, in order to improve our knowledge of the kinematics of the inner
luminous regions, deep IFS observations are mandatory, and they 
are already being performed on large sample of nearby 
galaxies by the SAURON team.

Here we discussed an observational project to survey the full 2D kinematics of 
intermediate luminosity/disky galaxies, up to about 1.2  $R_{\rm e}$, in order
to overlap the region where PNe data are available.
The preliminary sample includes galaxies which have been 
modeled with nearly constant mass-to-light ratios up to
about 6 $R_{\rm e}$ using the 
information from PNe discrete radial velocity fields (Romanowsky et al. 2003). 
This finding is extremely interesting, as it seems to violate the expectations 
from CDM hierarchical merger models (e.g., Navarro,
Frenk \& White 1997), which predict a substantial amount of dark matter in 
the outskirt of ellipticals.

As discussed above,
there are still some indeterminacies due to the intrinsic structure of 
disky systems (like the presence of faint disks or multicomponent with 
misaligned kinematics) and geometry (i.e., inclination):
the use of the detailed 2D kinematics and high velocity resolution 
is therefore necessary to 
clarify any presence of biases in the dynamical modeling due to 
kinematical substructures.

IFU-VIMOS is a very suitable instrument for such a purpose as it allows to 
combine an adequate spatial and spectral resolutions with a large survey field 
and the advantage of an 8m telescope collecting power.

\acknowledgements

NRN thanks S.C.Trager for useful discussion on S0s.
GC is being supported by the RTN Euro3D postdoctoral fellowship.
NRN is receiving grant from the European Community 
(FP5 Program -- Human Resources, Marie Curie Fellowship). 
The authors thank the anonymous referee for a careful reading 
of the paper and are grateful to the PN.S team for allowing the use 
of unpublished data.


\begin{thebibliography}{}
\bibitem{} Bacon, R. et al. 2001, MNRAS, 326, 23
\bibitem{} Bender, R., et al. 1988, A\&AS, 74, 385
\bibitem{} Bender, R., Saglia, R.P. \& Gerhard, O.E., 1994, MNRAS, 269, 785
\bibitem{} Bertin, G. et al. 1994, A\&A 292, 381
\bibitem{} Bertola, F. \& Capaccioli, M., 1975, ApJ, 200, 439
\bibitem{} Capaccioli, M., 1987, in IAU Symposium 127, 
{\em Structure and Dynamics of elliptical galaxies}, 
p. 47, ed. T. de Zeeuw, R. Dorthrect 
\bibitem{} Capaccioli, M., Caon, N., \& D'Onofrio, 
M. 1992, ESO ESP/EIPC Workshop on Structure of Early-type Galaxies, 
eds. J. Danziger, W. W. Zeilinger, and K. Kjar, ESO: Garching, 43
\bibitem{} Capaccioli, M., Napolitano, N. R., Arnaboldi, M. 2003, 
Sakharov Conference of Physics, Moskow, June 2002, in press, 
[preprint:astro-ph/0211323] 
\bibitem{} Davies,, R., et al. 1983, ApJ, 266, 41
\bibitem{} de Zeeuw, P.T. et al.: 2002, MNRAS 329, 513
\bibitem{} D'Odorico, S. et al 2003, The Messenger 113, 26 
\bibitem{} Douglas, N.G., et al. 2002, PASP, 114, 1234
\bibitem{} Faber, S.M. et al. 1997, AJ, 114, 1771
\bibitem{} Foucaud, S. et al.:2003 this volume
\bibitem{} Kissler-Patig, M., 1997, A\&A, 319, 83
\bibitem{} Lokas, E., 2002, MNRAS, 333, 697
\bibitem{} Magorrian \& Ballantyne 2001, MNRAS 322, 702
\bibitem{} M\'{e}ndez, R.H. et al. 2001, ApJ, 563, 135 
\bibitem{} Meza, A., et al., 2003, ApJ, 590, 619
\bibitem{} Milgrom, M. \& Sanders, R.H. 2003, preprint [astro-ph/0309617]
\bibitem{}  Napolitano, N.R., Arnaboldi, M., Freeman, K.C. 
\& Capaccioli, M. 2001, A\&A, 377, 784 
\bibitem{} Napolitano et al., 2003, on ASP proceedings of the IAU Symposium 220 "Dark matter in galaxies", 
eds. S.Ryder et al., in press, [preprint: astro-ph/0310798]
\bibitem{} Napolitano, N.R. 2001, \textit{Extragalactic 
Planetary Nebulae as tracers of the mass distribution in early-type 
galaxies and
clusters}, PhD Thesis, Universit\`a ``Federico II'', Naples (Italy) (Np01)
\bibitem{} Navarro,  J.F.,  Frenk, C.S. \& White,S.D.M. 1997, ApJ 490, 493 
\bibitem{} Nieto, J.L., \& Bender, R. 1989, A\&A, 215, 266 
\bibitem{} Pellegrini, S. 1999, A\&A, 351, 487 
\bibitem{} Pignatelli, E. \& Galletta, G., 1999, A\&A, 349, 369 
\bibitem{}  Romanowsky, A.J. \& Kochanek, C.S., 2001, ApJ, 553, 772
\bibitem{} Romanowsky, A.J. et al. 2003, Science, 301, 1696 [see also, astro-ph/0308518] (R+03)
\bibitem{} Scorza, C. \& Bender, R., 1995, ApJ, 293, 20
\bibitem{} Statler, T.S. \& Smecker-Hane, T.: 1999, AJ 117, 839 
\bibitem{} Verolme, E.K. et al. 2002, MNRAS 335, 517

\end{thebibliography}
\end{document}